\documentclass[preprint2]{aastex}

\shorttitle{Sub-systems in wide  binaries}
\shortauthors{Tokovinin, Hartung, Hayward}

\begin{document}

\title{Sub-systems in  nearby solar-type wide binaries\footnote{Based
    on observations obtained at the Gemini Observatory (Program ID
    GS-2009B-Q-49), which is operated by the Association of
    Universities for Research in Astronomy, Inc., under a cooperative
    agreement with the NSF on behalf of the Gemini partnership: the
    National Science Foundation (United States), the Science and
    Technology Facilities Council (United Kingdom), the National
    Research Council (Canada), CONICYT (Chile), the Australian
    Research Council (Australia), Minist\'{e}rio da Ci\^{e}ncia e
    Tecnologia (Brazil) and Ministerio de Ciencia, Tecnolog\'{i}a e
    Innovaci\'{o}n Productiva (Argentina).}
}

\author{Andrei Tokovinin}
\affil{Cerro Tololo Inter-American Observatory, Casilla 603, La Serena, Chile}
\email{atokovinin@ctio.noao.edu}

\author{Markus Hartung, and Thomas L. Hayward }
\affil{Gemini Observatory, Southern Operations Center, c/o AURA, Casilla 603, La Serena, Chile}
\email{mhartung@gemini.edu, thayward@gemini.edu}


\begin{abstract}
We conducted a deep survey of resolved sub-systems among wide binaries
with solar-type components within 67 pc from the Sun.  Images of 61
stars in the $K$ and $H$ bands were obtained with the NICI adaptive-optics
instrument on the 8-m Gemini-South telescope.  Our maximum detectable
magnitude difference is about $5^m$ and $7.8^m$ at $0.15''$ and
$0.9''$ separations, respectively. This enables a complete census of
sub-systems with stellar companions in the projected separation range
from 5 to 100 AU.  Out of 7 such companions found in our sample, only
one was known previously. We determine that the fraction of
sub-systems with projected separations above 5 AU is $0.12 \pm 0.04$
and that the distribution of their mass ratio is flat, with a
power-law index $0.2 \pm 0.5$.  Comparing this with the properties of
closer spectroscopic sub-systems (separations below 1~AU), it appears
that the mass-ratio distribution does not depend on the separation.
The frequency of sub-systems in the separation ranges below 1\,AU and
between  5 and 100\,AU is similar, about 0.15. Unbiased statistics of
multiplicity higher than two, advanced by this work,  provide
constraints on  star-formation theory. 
\end{abstract}


\section{Introduction}
\label{sec:intro}

The quest for exo-planets has attracted attention to stars in the
solar neighbourhood.  A complete understanding of their formation
requires a quantitative and predictive description of stellar
multiplicity.  Empirical data on stellar multiplicity serve as a
benchmark for star formation theories, for example to compare with
recent large-scale hydrodynamical simulations \citep{Bate09}.
Although multiplicity statistics of solar-type dwarfs were
assumed to be known since the survey of \citet{DM91}, in fact that
survey did not constrain multiplicities higher than two because (i)
the sample of 164 stars was too small and (ii) the discovery of wide
binaries relied on traditional visual techniques.  \citet{Raghavan09}
shows that \citet{DM91} missed half of higher-order multiples within
25\,pc from the Sun.

Several studies of large samples of nearby stars are now in progress.
Radial-velocity  (RV)  techniques ensure  a  secure  detection of  all
short-period  stellar companions \citep[see  summary in][]{Grether06}.
However, for wider separations ($>$ 10\,AU) RV variations are too slow
or too small, and such companions  are better discovered by direct imaging.
Several recent deep imaging surveys, summarized by \citet{MH09}, aimed
to  detect substellar  or planetary-mass  companions to  nearby stars,
obtaining statistics of {\em stellar} companions as a by-product.  The
typical sample size of imaging surveys is still small, on the order of
100 targets.

Most imaging  and RV surveys for low-mass  companions explicitly avoid
known   visual   binaries   \citep[e.g.][]{Chauvin10},  so   important
star-formation effects  may not be represented in  their samples.  For
example, catalogued triples show  that the mass-ratio distributions in
close  inner  sub-systems discovered  spectroscopically  and in  wider
``visual'' sub-systems are  radically different (see Section~5).
One of our goals is to investigate  if this is a selection effect or a
genuine feature of the star-formation process.

In order to improve the  multiplicity statistics of a well-defined and
sufficiently  large distance-limited  sample, we  focus  on $\sim$5000
dwarfs with $0.5  < V-I < 0.8$ within 67\,pc  from the Sun.  According
to preliminary  estimates, there  should be some  400 triples  and 100
quadruples  in this  sample.  Surveying  such a large sample to
detect those  multiples with a reasonably  deep  and known completeness
limit  is  a  large task;   here  we  address  only  a small  part  of
it. Instead  of avoiding known binaries, we  purposefuly selected wide
visual  binaries from  this sample.   Some of  these targets  are also
monitored in RV, giving  complementary constraints on sub-systems with
short periods \citep{Desidera06}.

We imaged the chosen wide binaries  with Adaptive Optics (AO)
to study the properties of inner sub-systems.  AO imaging on 8-m class
telescopes provides nearly diffraction-limited resolution of 0.06$''$
in the near-infrared, which corresponds to a projected orbital radius
of 4\,AU for the farthest target stars in our sample, filling an
important gap between RV and seeing-limited imaging surveys.

Our sample is detailed in Section~\ref{sec:sample}. Observations, data
reduction,     and    detection     limits     are    described     in
Section~\ref{sec:obs},      the     results      are      presented    in
Section~\ref{sec:res}. Sections~\ref{sec:disc} and \ref{sec:concl}
contain the discussion and conclusions, respectively. 

\section{The sample of wide binaries}
\label{sec:sample}

The wide binaries chosen for our survey 
belong to a large sample of nearby
solar-type dwarfs ({\it Nsample}) selected from the latest version of Hipparcos
catalog \citep{vLHIP} by the following criteria:
\begin{enumerate}
\item
trigonometric parallax $\pi_{\rm HIP}  < 15$\,mas (within 67\,pc from
the Sun, distance modulus  $<4.12^m$);

\item
color $0.5 < V-I < 0.8$ (this corresponds approximately to spectral types from F5V to K0V);

\item
unevolved, satisfying the condition  $M_{\rm Hp} > 9(V-I) - 3.5$,
where $M_{\rm Hp}$ is the absolute magnitude in the Hipparcos band
calculated with  $\pi_{\rm HIP}$. 

\end{enumerate}
There  are 5040  stars  (i.e.  Hipparcos  catalog entries)  satisfying
these  conditions. Some of  these stars  with erroneous  parallaxes or
colors  will   be  eventually  removed  from  the   sample,  but  this
contamination  is expected  to be  minor.  The  {\it  Nsample} defined
above  is  reasonably complete  because  the  cumulative object  count
versus distance $d$  follows the $d^3$ law, deviating  from it by only
10\% at $d = 67$\,pc.

Visual binaries with  separations $5'' < \rho < 20''$  and one or both
components  belonging  to  the  {\it  Nsample} were  selected  for  AO
observations from the Washington Double-Star Catalog, WDS \citep{WDS}.
Additionally, we require each pair to have at least 3 observations and
magnitudes of  both components  listed in the  WDS, and  that apparent
motion of the  wide pair be compatible with  $\pi_{\rm HIP}$.  As most
targets have  large proper motions,  relative stability of  the binary
position  over time permits  rejection of  optical pairs.   An additional
check  was  made  by  placing  the  components  on  the  $(M_V,  V-K)$
color-magnitude diagram  (CMD).  There are $\sim$200  wide binaries in
this list,  101 with negative declinations.

\begin{deluxetable}{l ccc ccc ccc c c }            
\tabletypesize{\scriptsize}                                                                                                               
\tablecaption{Data on observed targets
\label{tab:tar}   }                                                                                                                       
\tablewidth{0pt}                                                                                                                          
\tablehead{  HIP (A/B)   & $\pi_{\rm HIP}$ &  $\rho$ & $r$ &  \multicolumn{3}{c}{Component A} &  \multicolumn{3}{c}{Component B} & Date & Rem \\ 
               &   mas         &   $''$   & AU &   $V$ & $H$ & $K_s$ &   $V$
  & $H$ & $K_s$ & 2000+  & }
\startdata                                                                                                              
 2028/29        &  24.0  &  16.5 & 687 &   8.39  &  7.46  &  7.41 &  9.56  &  8.30  &  8.27 &  09.9400& B \\ 
10579           &  17.2  &   6.7 & 390 &   9.43  &  8.06  &  7.97 &  9.97  &  8.34  &  8.24 &  09.9400& \\
20552           &  36.1  &   5.6 & 155 &   6.87  &  5.37  &  5.32 &  7.23  &  5.64  &  5.55 &  10.0851& \\
20610/12        &  16.7  &  10.5 & 629 &   8.06  &  6.89  &  6.84 &  8.27  &  7.01  &  6.98 &  10.0851& B \\
21963           &  18.2  &   8.2 & 450 &   8.13  &  6.69  &  6.62 & 13.30  &  8.99  &  8.75 &  10.0852& \\
22531/34        &  26.0  &  12.9 & 496 &   5.61  &  4.88  &  4.80 &  6.24  &  5.28  &  5.19 &  09.9404& A,CRN \\
23926/23        &  18.7  &  10.1 & 540 &   6.83  &  5.20  &  5.10 & 10.29  &  8.00  &  7.90 &  10.0851& \\
24711/12        &  15.3  &  13.3 & 867 &   8.46  &  7.00  &  6.95 & 10.57  &  8.29  &  8.24 &  10.0852& \\
27922           &  42.4  &  10.4 & 246 &   7.60  &  5.84  &  5.76 & 10.57  &  7.40  &  7.22 &  10.0852& \\
28790           &  37.2  &   5.6 & 150 &   6.02  &  4.92  &  4.75 &  8.98  &  6.09  &  6.03 &  10.0853& \\
30158           &  17.8  &   7.0 & 390 &   8.57  &  7.02  &  6.91 & 10.58  &  8.09  &  8.03 &  10.0853& \\
32644           &  21.2  &   4.9 & 230 &   7.37  &  6.05  &  5.97 &  8.66  &  6.61  &  6.52 &  10.0852& \\
36165/60        &  31.2  &  17.7 & 566 &   7.08  &  5.92  &  5.80 &  8.06  &  6.75  &  6.60 &  10.0854& \\
37332/33        &  15.1  &  15.7 &1043 &   7.89  &  6.47  &  6.41 & 10.01  &  8.35  &  8.25 &  10.0855& \\
37335           &  21.8  &   6.1 & 282 &   8.93  &  7.05  &  6.99 & 12.38  &  8.59  &  8.40 &  10.0855& \\
39409           &  15.7  &   5.1 & 328 &   9.26  &  7.57  &  7.44 &  9.36  &  7.56  &  7.43 &  10.0855& \\
43947           &  26.6  &  10.2 & 383 &   8.44  &  7.28  &  7.19 &  9.95  &  8.27  &  8.20 &  10.0857& \\
44584/85        &  17.0  &  10.5 & 617 &   8.09  &  6.99  &  6.90 &  9.19  &  7.72  &  7.70 &  10.0857& B \\
44804           &  15.5  &   7.4 & 478 &   8.80  &  7.20  &  7.15 & 10.70  &  8.45  &  8.29 &  10.0857& \\
45734           &  17.6  &   9.0 & 513 &   8.41  &  6.85  &  6.78 &  9.66  &  7.56  &  7.44 &  10.0856& \\
45940           &  28.6  &   6.8 & 237 &   8.17  &  6.55  &  6.43 & 11.86  &  8.35  &  8.07 &  10.0856& \\
46236           &  21.4  &  19.3 & 900 &   7.03  &  5.77  &  5.69 & 10.18  &  7.82  &  7.68 &  10.0856& \\
47839/36        &  40.2  &  18.8 & 468 &   8.08  &  6.79  &  6.70 &  8.23  &  6.99  &  6.85 &  10.0858& \\
49520           &  16.9  &   9.6 & 565 &   8. 82  &  7.35  &  7.24 &  8.99  &  7.40  &  7.30 &  10.0858& \\
50638/36        &  83.3  &  17.1 & 205 &   7.67  &  5.18  &  4.98 &  9.67  &  8.46  &  8.40 &  10.0858& \\
50883           &  16.2  &   6.7 & 413 &   7.90  &  6.71  &  6.62 & 13.00  &  9.05  &  8.98 &  10.0858& \\
55288           &  20.8  &   9.5 & 456 &   7.10  &  5.91  &  5.82 &  7.91  &  6.32  &  6.22 &  10.0859& \\
58240/41        &  21.1  &  18.9 & 895 &   7.67  &  6.24  &  6.13 &  7.83  &  6.29  &  6.24 &  10.0859& \\
59021           &  19.4  &   6.0 & 308 &   6.69  &  5.24  &  5.10 &  8.84  &  6.71  &  6.55 &  10.0859& CRN \\
61595           &  17.8  &   8.4 & 472 &   8.28  &  6.80  &  6.68 & 10.48  &  8.29  &  8.12 &  10.0860& A \\
64498           &  18.0  &   9.3 & 516 &   7.74  &  6.40  &  6.25 & 10.10  &  7.98  &  7.85 &  10.0861& \\
65176           &  17.8  &   5.0 & 282 &   7.85  &  6.47  &  6.40 &  8.59  &  6.82  &  6.68 &  10.0861& \\
67408           &  33.9  &  11.6 & 342 &   6.62  &  5.29  &  5.22 & 10.21  &  7.14  &  7.02 &  10.0860& A,CRN\\
\enddata                                                                                                                                  
\end{deluxetable}

In Table~\ref{tab:tar} we list basic  data on the components of the 33 wide
pairs observed  in this run. Column  1 gives the  Hipparcos numbers of
the primary and, in some cases, secondary components. The next three columns
contain $\pi_{\rm HIP}$, binary separation $\rho$, and projected
separation $r = \rho/\pi_{\rm HIP}$.   Then follow the
magnitudes of the primary (A) and secondary (B) components in the $V$,
$H$,  and  $K_s$  bands as  listed  in  the  WDS and  2MASS
\citep{2MASS}. The  last two columns contain the  date of observations
and remarks where we indicate 5 cases when only one component
was observed and 3 cases when the coronagraphic mask was used.

\section{Observations and data reduction}
\label{sec:obs}

\subsection{Observing procedure}

The  Near-Infrared Coronagraphic  Imager,  NICI, on  the Gemini  South
telescope is  an 85-element curvature  AO instrument based  on natural
guide   stars  \citep{NICI,   Chun08}.    We  used   NICI  in   normal
(non-coronagraphic)  mode,   as  a  classical   AO  system.   However,
simultaneous imaging in the two  infrared bands $H$ and $K$ offered by
NICI helps  to distinguish faint companions from  static speckles (the
radial distance of  a static speckle scales with  wavelength while the
position of a real detection  does not change). The two detectors have
$1024^2$ pixels  of 18\,mas (milliarcseconds) size,  covering a square
field of $18''$.

To  avoid  saturation,  we   used  narrow-band  filters  with  central
wavelengths  2.272\,$\mu$m  and  1.587\,$\mu$m  in the  red  and  blue
channels,  respectively.   The   minimum  possible  exposure  time  is
0.38\,s,  still causing  saturation for  bright targets.   We observed
three of the  brightest stars with the coronagraphic  mask of $0.32''$
radius, using in this instance  broadband $K$ and $H$ filters. For the
remaining bright targets we allowed moderate saturation. As the tradeoff
between slower observations with more complex data reduction in
coronagraphic mode and slightly saturated PSFs is not obvious, we tried
both. 

Typically, each target  was observed at 5--6 dithered  positions for a
total  accumulation time  of about  5\,min.  The  dither  offsets were
designed to take  images of both wide-binary components  in one frame
whenever possible  ($\rho <  10''$) while closing  the AO loop  on the
primary. Components  of wider  pairs were observed  separately, except
secondaries with $V > 12^m$, too faint to serve as AO guide stars.

Of the  two nights allocated  to this program,  one was lost due  to a
technical problem.  Three targets were observed on 2009 December 10 in
service mode.   The remaining  objects were observed  on the  night of
2010 January  31 to February 1  in classical mode.   The seeing during
that night was good. As measured from the NICI AO loop data, it ranged
between $0.34''$ and $0.94''$, with  a median of $0.50''$ (at 500\,nm,
not corrected to  the zenith).  High AO performance  was achieved (see
next Section).  The transparency  was variable (light cirrus).  Out of
202 binary components accessible to  NICI, we were able to observe 61,
or  30\%.  The efficiency  was high,  with an  average of  12\,min per
target  including  telescope  slew,  tuning  of  the  primary  mirror,
acquisition of the object on  the science camera, closing the AO loop,
and data-taking.

\subsection{Data processing}

The data were  processed in a standard way  using custom IDL programs.
Individual images  written to separate  FITS files were  combined into
data cubes,  separately in  red and blue  channels. The sky  image was
obtained for each  target as a median of the  cube, then subtracted
from  each plane of  the cube.   After correcting  for bad  pixels and
dividing  by   the  flat  field,   the  images  were   re-centered  by
cross-correlation and median-combined.

\begin{figure*}[ht]
\epsscale{2.0}
\plotone{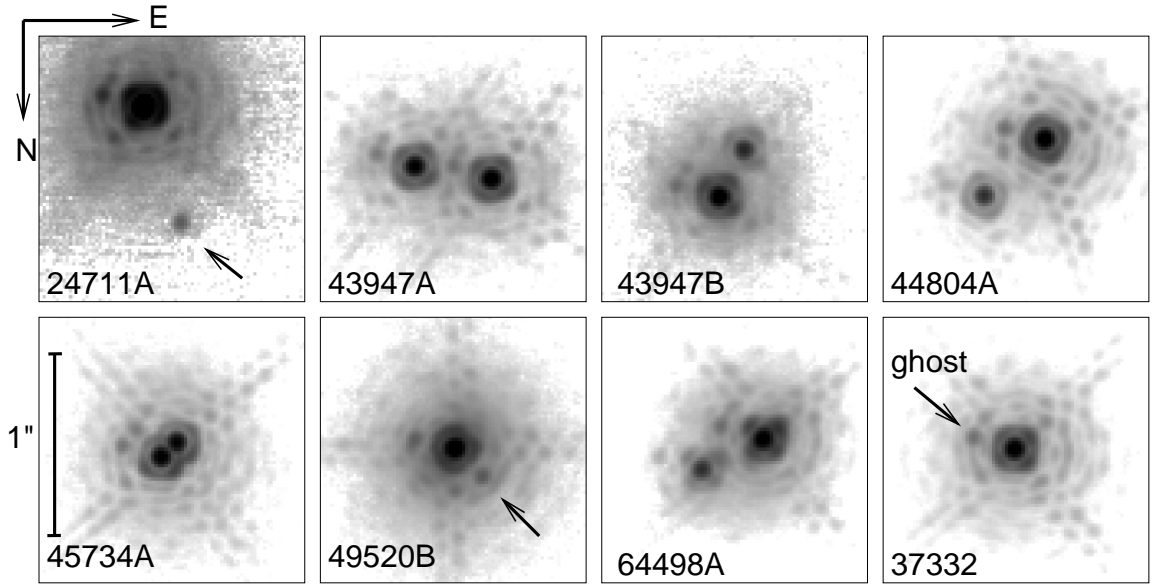}
\caption{Mosaic of resolved-component images in the r-channel, labeled
  by  the  Hipparcos numbers.   Each  frame  is  81 pixels  ($1.46''$)
  across, oriented  North down and  East to the right.   The intensity
  scaling is negative logarithmic.  
The first 7 frames show all binary components where we detected a close
companion. 
The  last frame shows the PSF of the
  unresolved  target  HIP~37332,  with  the ghost  marked.  The  faint
  companions to HIP~24711A and  HIP~49520B are marked by arrows.  Note
  that  at good AO  correction the  typical features  of the  NICI PSF
  become visible. These are particularly the double diffraction spikes
  and the  four spots  on the  first Airy ring,  giving it  a ``boxy''
  look. These features  are caused by the oversized  vanes of a spider
  mask placed in the pupil plane to control scatter.
\label{fig:images} 
}
\end{figure*}

Most frames contain just one or two bright point sources corresponding
to the components of the wide binary.  
  The position  of  each point
source  was recorded  by ``clicking''  on  its image,  then refined  by
fitting  a 2-dimensional  Gaussian to  the central  part of  the Point
Spread Function (PSF) within 5-pixel radius.  The ratio of the central
intensity of  this Gaussian  to the total  flux in a  $0.9''$ diameter
circle  gives   an  estimate  of   the  Strehl  ratio.    The  typical
Strehls\footnote{The oversized spider mask has been taken into account
  and yields  a correction factor of  1.079} are 43\% and  22\% in the
red and blue channel,  respectively, with maximum values reaching 54\%
and 38\%. The median Full Width at Half Maximum (FWHM) of the PSF core
was   65\,mas   and   57\,mas   in   the  red   and   blue   channels,
respectively. The diffraction limit $\lambda/D$ is 63\,mas and 44\,mas
for   the  effective   aperture   diameter  $D=7.5$\,m   set  by   the
coronagraphic mask.

Figure~\ref{fig:images} shows the mosaic  of 7 close sub-systems found
around some primary components  of wide binaries.  Most companions are
rather obvious, well above  the detection threshold.  The PSFs contain
also  a  faint  secondary  reflex  at 13  pixels  ($0.24''$),  with  a
magnitude difference in  the red channel of about  $4.3^m$.  This {\em
  ghost}  (illustrated   in  Fig.~\ref{fig:images}  in   the  case  of
HIP~37332) is detected  in all targets, and can  be taken advantage of
as a  fiducial (see  Section 3.4).  Its position is  the same  in both
channels, whereas  the static speckle pattern scales  in proportion to
the wavelength.

\subsection{Measurement of binaries}

To measure the relative position  and flux ratio of wide binaries more
precisely,  we  fit  the  scaled  and shifted  image  of  the  primary
component (A) to the secondary  (B). Such fits work very well, leaving
only small residuals, and produce mutually consistent measurements in
the  two  channels.  The  rms scatter  between  the wide-binary  positions
measured in two NICI channels is 5.5\,mas in both radial and tangential
directions.

We checked the  pixel scale and detector orientation  by comparing the
measured positions of 16 wide pairs (the widest in Table~\ref{tab:res}
after exclusion  of HIP~21963, 43947  and 50883, which were  not used)
with the  positions listed in  the Hipparcos Double and  Multiple Star
Catalogue  \citep{HIP1997d}.  We  did not  find any  systematic effect
above noise  and therefore used  the nominal pixel scale  of 18.0\,mas
for  both channels.   We found  that the  detector orientation  of the
channels differs by $1.01^\circ \pm 0.01^\circ $, so $1.0^\circ $ were
added to measured angles in the blue channel.

For close  pairs where the  PSFs overlap, we  use a variant  of iterative
blind  deconvolution.   Good  preliminary  estimates of  the  relative
component's  position and  intensity  are already  available from  the
Gaussian fits of  the PSF cores.  The image  Fourier Transform (FT) is
divided by the FT of the  binary with preliminary parameters to get an
estimate of the PSF. Negative parts of this PSF are set to zero and it
is replaced by its azimuthal average at radial distances over 5 pixels
from the center.   The binary parameters are refined  by fitting again
the  image with  this synthetic  PSF.  Then  the process  is repeated,
obtaining a second approximation to  the PSF and using it again (without
azimuthal  averaging)  to  obtain  the final  binary  parameters.  The
success of this procedure is verified by inspecting the resulting PSF,
which  should be  ``clean,''  with  the least  possible  trace of  the
companion.

\subsection{Detection limits}

\begin{figure}[ht]
\epsscale{1.0}
\plotone{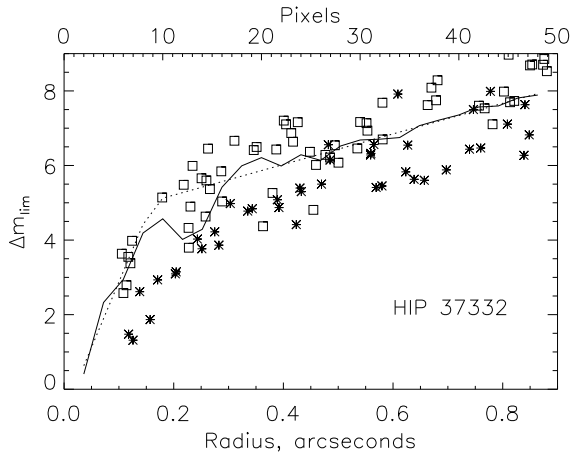}
\caption{Estimated $5 \sigma$ companion detection limit (solid line)
  in  the red channel  for HIP\,37332.  The simulated  companions are
  over-plotted as asterisks (detected) or empty squares (missed). The
  dotted line shows the 2-segment model.
\label{fig:simdet} 
}
\end{figure}

To determine detection limits, we use mostly the data from the red
NICI channel where the detection of low-mass components is deeper
because of the smaller magnitude difference and larger Strehl ratios.
The maximum magnitude difference of detectable companions $\Delta m_{\rm
  lim}$ is estimated by calculating the rms flux variations
$\sigma(\rho)$ in circular zones of increasing radii $\rho$ centered
on the main component. It is assumed that a detectable companion will
have maximum intensity exceeding $5 \sigma$, therefore
\begin{equation}
\Delta m_{\rm lim} (\rho) <  -2.5 \log_{10} [ 5 \sigma(\rho) / I_{\rm max}
] .
\label{eq:dm}
\end{equation}
This assumption is,  of course, simplistic. We do  not account for the
existence of two simultaneous images  in red and blue channels and for
the good knowledge of the PSF from other companions of the same target
or from other targets observed shortly before or after.  Nevertheless,
this simple strategy of  estimating detection limits is sufficient for
our   purpose  and   has   been  tried   before   with  good   results
\citep[e.g.][]{Tok06}.

The limits computed for  HIP~37322 are shown in Fig.~\ref{fig:simdet}.
A bump  is produced in the curves  at $\rho \approx 13$  pixels by the
ghost   and    by   fixed   speckle   pattern   (see    the   PSF   in
Fig.~\ref{fig:images}).  To confirm  the validity of (\ref{eq:dm}), we
performed Monte-Carlo simulations, adding artificial companions within
$\pm 1.5^m$ from the  estimated $\Delta m_{\rm lim}$. Secure detection
of the  ghost ($\Delta m =  4.3$, $\rho =0.24''$) in  all images gives
confidence in our estimation of $\Delta m_{\rm lim}$.

The  $5 \sigma$  detection  limits $\Delta  m_{\rm  lim} (\rho)$  were
fitted  by  two straight  lines  intersecting  at  $\rho =  8$  pixels
($0.144''$). The second segment extends from 8 to 50 pixels ($0.9''$),
but the region between 8 and 16 pixel radius affected by the ghost and
strong speckle is avoided in  the fit.  At $\rho > 50$\,pixels $\Delta
m_{\rm  lim}$ is  assumed to  be constant.   Only two  numbers $\Delta
m(8)$ and $\Delta m(50)$ adequately describe the actual $\Delta m_{\rm
  lim}  (\rho)$  curves,  with  typical  rms  error  of  only  $0.3^m$
(excluding the zone between 8 and  16 pixel radii where the $5 \sigma$
method does not properly account for the static speckle).

Detection limits were determined by the above procedure for all
targets, although the Monte Carlo simulation checks were only performed
for a subset.
The  two detection parameters  $\Delta m(8)$  and $\Delta  m(50)$ vary
little  between  targets.  Their  median  values  are  5.02 and  7.79,
respectively; the quartiles are (4.29,  5.22) and (7.03, 8.06). In the
following  we  apply  median  detection  parameters  to  all  targets,
including  the  three coronagraphic  images  where  the actual  limits
should be deeper.

The detection limits found here  match other similar studies done with
``standard'' AO.   For example, companion detection  was complete down
to $\Delta  K < 3.9$ at  $0.14''$ and $\Delta  K < 7.9$ at  $0.9''$ in
\citep{Tok06}. Similar results were obtained by \citet{Eggenberger07}.
The  deep   companion  survey  at   Palomar  and  Keck   conducted  by
\citet{MH09}  also reached  $\Delta  K  < 8$  at  $0.9''$.  Note  that
typical  detection limits  with  NICI are  much  deeper for  dedicated
planet  search  programs.  In  these  cases long  exposure  times  and
techniques  such  as angular  and  spectral  differential imaging  are
applied \citep{Biller08,Artigaut08}.

\section{Results}
\label{sec:res}

\begin{deluxetable}{c c  ccc ccc c }
\tabletypesize{\scriptsize}                                                                                                               
\tablecaption{Measured parameters of known and new pairs
\label{tab:res} }
\tablewidth{0pt}                                                                                                                          
\tablehead{  HIP    & Comp   & \multicolumn{3}{c}{Red channel (2.272\,$\mu$m)} &
  \multicolumn{3}{c}{Blue  channel (1.587\,$\mu$m) } & Rem \\
       &        & $\theta$  & $\rho$ & $\Delta m$ & $\theta$  & $\rho$ & $\Delta m$ & \\ 
       &        & $^\circ$   & $''$   &   mag      & $^\circ$   & $''$
  &   mag      & }    
\startdata     
10579  &  A,B  &   288.58 &  6.710 &   0.47  &  288.50 &  6.703  &  0.71&   \\
20552  &  A,B  &   247.35 &  5.405 &   0.16  &  247.55 &  5.403  &  0.26& sat?  \\
21963  &  A,B  &    92.33 &  8.184 &   2.35  &   92.26 &  8.169  &  2.64&   \\
23926  &  A,B  &   259.93 & 10.255 &   2.35  &  259.96 & 10.246  &  2.66& sat  \\
24711  & Aa,Ab &    17.70 &  0.670 &   5.28  &   17.71 &  0.663  &  5.66& new  \\
24711  &  B,C  &   324.85 & 11.135 &   6.29  &  324.87 & 11.132  &  6.45& opt  \\
27922  &  A,B  &    19.35 & 10.742 &   1.55  &   19.37 & 10.727  &  1.88&   \\
28790  &  A,B  &   215.28 &  5.945 &   1.29  &  215.18 &  5.941  &  1.48& sat  \\
30158  &  A,B  &     7.48 &  7.096 &   1.18  &    7.38 &  7.093  &  1.33&   \\
32644  &  A,B  &   160.49 &  5.029 &   0.72  &  160.43 &  5.039  &  0.73& sat?  \\
37735  &  A,B  &   128.43 &  6.210 &   1.57  &  128.49 &  6.200  &  1.76&   \\
39409  &  A,B  &   329.64 &  5.187 &   0.10  &  329.67 &  5.193  &  0.14&   \\
43652  & A,Ca  &   128.47 &  9.206 &   6.16  &  128.48 &  9.188  &  6.41& opt  \\
43652  &Ca,Cb  &   328.86 &  0.124 &   0.67  &  330.90 &  0.126  &  0.64& opt  \\
43947  &Aa,Ab  &   260.79 &  0.424 &   0.31  &  260.78 &  0.423  &  0.30& new  \\
43947  &Ba,Bb  &   152.16 &  0.289 &   1.65  &  152.16 &  0.289  &  1.73& new  \\
43947  &Aa,Ba  &   202.89 & 10.377 &   0.82  &  202.78 & 10.365  &  0.90&  \\
44804  & Aa,B  &   136.01 &  7.175 &   1.16  &  136.08 &  7.167  &  1.21&   \\
44804  &Aa,Ab  &   313.09 &  0.453 &   1.84  &  313.09 &  0.452  &  2.02& new  \\
45734  &Aa,Ab  &   132.69 &  0.119 &   0.38  &  133.33 &  0.117  &  0.36&   \\
45734  & Aa,B  &   193.63 &  9.077 &   0.71  &  193.65 &  9.076  &  1.23&   \\
45940  &  A,B  &    69.28 &  6.713 &   1.72  &   69.29 &  6.705  &  1.98&   \\
47836  &  A,C  &   328.29 &  9.809 &   5.31  &  328.36 &  9.797  &  5.77& opt  \\
49520  &  A,B  &   326.95 &  9.552 &   0.29  &  326.95 &  9.543  &  0.44&   \\
49520  &Ba,Bb  &    44.21 &  0.213 &   3.61  &   46.69 &  0.210  &  3.40& new  \\
50883  &  A,B  &   192.65 &  6.839 &   2.33  &  192.57 &  6.830  &  2.57&   \\
50883  &  A,C  &   299.87 &  7.379 &   6.23  &  299.82 &  7.363  &  6.55& opt  \\
55288  &  A,B  &   253.94 &  9.462 &   0.54  &  253.96 &  9.456  &  0.84&   \\
58241  &  A,C  &   118.32 &  4.015 &   6.38  &  118.09 &  4.008  &  6.52&   \\
64498  & Aa,B  &   302.20 &  9.464 &   1.59  &  302.24 &  9.458  &  1.91&   \\
64498  &Aa,Ab  &   296.55 &  0.372 &   1.76  &  296.31 &  0.370  &  2.01& new  \\
65176  &  A,B  &    96.89 &  5.168 &   0.61  &   96.95 &  5.164  &  0.62&   \\
\enddata                                                                                                                                  
\end{deluxetable}

\subsection{Companion data}

We have found 6 new  close companions, one previously known companion,
and   several   faint   optical  companions   (Fig.~\ref{fig:images}).
Table~\ref{tab:res}  lists   the  relative  positions   and  magnitude
differences in known and new pairs. Each pair is identified by the HIP
number  and component  designations. The  measured position  angles in
degrees,  separations  in arcseconds,  and  magnitude differences  are
listed separately for the two NICI channels.

The large rms difference between Hipparcos and our measurements of the
wide pairs (130\,mas in $\rho$ and $1.05^\circ$ in $\theta$) is mostly
caused by  their motion between  the two epochs.  The  comparison with
the relative  positions of  the same binaries  deduced from  the 2MASS
Point Source  Catalog \citep{2MASS}  has confirmed that  no systematic
corrections are needed and has shown a scatter of similar magnitude.

Remarks in the last  column of Table~\ref{tab:res} indicate cases when
the image was saturated  (the relative photometry may be compromised),
newly detected close companions,  and likely optical companions. These
faint  companions  with  separations  of  a few  arcseconds  would  be
dynamically   unstable   within   wide   binaries,  should   they   be
physical\footnote{Dynamical stability  of a triple  system is possible
  when the ratio of the  inner semi-major axis to the outer periastron
  distance  is larger  than  $\sim$3 \citep[e.g.][]{Harrington72}.  As
  only projected separations are  known for our triples, assessment of
  their dynamical stability is approximate.}.

\subsection{Notes on individual systems}

{\bf HIP~28790} has a common-proper-motion companion C at $196''$.

{\bf HIP 36165 A} is suspected to be a spectroscopic binary (SB). 

{\bf HIP 43947} is unexpectedly  discovered to be a resolved quadruple
system. Both components are located $\sim 2^m$ below the Main Sequence
in  the $(M_V,  V-K)$  CMD, possibly  because  the Hipparcos  parallax
measurement was biased by the motions in the sub-systems. The position
and $\Delta  m$ of the  wide pair Aa,Ba  is derived from  the Gaussian
fits, therefore it may be less accurate.

{\bf HIP  45734:} This is  a young pre-Main-sequence system  and X-ray
source RX  J0919.4-7738.  It  projects on the  Chamaeleon star-forming
region, but is located much closer than the corresponding association.
\citet{Covino97}  found from a  single spectrum  that   component B
(southern)  is a double-lined  SB,  with a  large RV
difference from  A \citep[see also][]{Desidera06}.   A  itself is a
close visual binary KOH83 resolved  in March 1996 by \citet{Koh01}. He
measured the position $(173.9^\circ , 0.109'')$ and $\Delta K = 0.45$.
This  is  the only  previously  known  resolved  sub-system among  the
observed  stars. Follow-up observations  will soon  lead to  the visual
orbit  and  mass  measurement,   adding  another  empirical  point  to  the
evolutionary tracks of young stars.

{\bf  HIP 49520:} The  faint companion  Bb is  close to  the detection
limit (see  Fig.~\ref{fig:images}).  Yet,  it is securely  detected in
both  channels and  stands up  clearly in  residuals when  fitting the
image of B with the image of A.

{\bf  HIP 50638/36:}  The  wide pair  has been  observed since  1835; it  is
definitely  physical. Component  B is  variable. For  this reason,
probably, it is located on the CMD some $6^m$ below the MS, while A is
on the  MS with the  parallax $\pi_{\rm HIP}  = 83$\,mas   listed in
Table~\ref{tab:tar}.  According  to \citet{vLHIP}, the  parallaxes and
proper motions  of the two  components are discordant, but  have large
errors. Future RV measurements can settle the controversy.

{\bf  HIP  59021:} Component  A is a  suspected  SB  and also  an
astrometric binary (acceleration solution in Hipparcos). Non-detection
of the sub-system with NICI means  that its period is likely less than
10\,y, or it is a white dwarf.

\subsection{Estimation of the mass ratios}

\begin{figure}[ht]
\epsscale{1.0}
\plotone{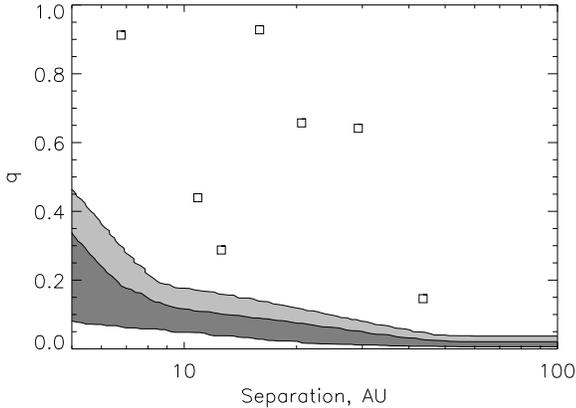}
\caption{Mass ratios $  q = M_2/M_1$ of 7 the resolved sub-systems versus
  their projected separations.  The shaded region indicates incomplete
  detections (light  grey corresponds to  detection probabilities from
  0.5 to 0.9, dark grey from 0.1 to 0.5).
\label{fig:q-sep} 
}
\end{figure}

$K$-band absolute magnitudes  $M_K$ of all  companions    were
computed based on the photometry from 2MASS (see Table~\ref{tab:tar}) and
trigonometric parallaxes $\pi_{\rm HIP}$ from Hipparcos.  For resolved
pairs,  we  assume  that  the  $\Delta  m$ in  the  red  NICI  channel
correspond to magnitude differences  in the $K$-band.  The masses were
estimated from $M_K$ using the empirical relations of \citet{HM93}. We
also  tried a  quadratic approximation  of relation  between  mass and
$M_K$   from  \citet{Girardi2000}.    The  relative   difference  with
\citet{HM93} is  less than 10\%  for masses above  0.1\,$M_\odot$.  We
checked that the  influence of using either of  these relations on the
final  result   is  not  critical.    Table~\ref{tab:mass}  lists  the
estimated  masses of  the companions  in the  resolved sub-systems,  their
projected   separations    $r   =   \rho   /\pi_{\rm    HIP}   $   and
order-of-magnitude   period   estimates   $P^*   =  r^{3/2}   (M_1   +
M_1)^{-1/2}$.

\begin{table}
\center
\caption{Estimated  parameters
of the 7 observed sub-systems in wide binaries}
\label{tab:mass}
\begin{tabular}{l c cc cc}
\tableline
HIP & Comp & $M_1$ & $M_2$ & r & $P^*$ \\
    &      & $M_\odot$ &  $M_\odot$ & AU & yr \\
\tableline
   24711 &   Aa,Ab &  1.04 & 0.15 &   43.6 &  260 \\
   43947 &   Aa,Ab &  0.64 & 0.59 &   15.9 &   60 \\
   43947 &   Ba,Bb &  0.55 & 0.24 &   10.9 &   40 \\
   44804 &   Aa,Ab &  0.95 & 0.61 &   29.3 &  130 \\
   45734 &   Aa,Ab &  0.88 & 0.81 &    6.8 &   14 \\
   49520 &   Ba,Bb &  0.92 & 0.26 &   12.6 &   40 \\
   64498 &   Aa,Ab &  1.09 & 0.71 &   20.6 &   70 \\
\tableline
\end{tabular}
\end{table}

Figure~\ref{fig:q-sep}  shows  the mass  ratios  $q=  M_2/M_1$ in  the
resolved inner sub-systems as a function of projected separation.  The
median parallax of observed targets  is 19\,mas, so the upper limit of
the separation  range, 100\,AU,  corresponds to $\rho  \approx 1.9''$.
Maximum  separations  of  dynamically  stable sub-systems  should  not
exceed 100\,AU in most cases  because the projected separations of the
wide      binaries      range      from      150      to      1000\,AU
(Table~\ref{tab:tar}).  Indeed,  all detected  sub-systems  have $r  <
44$\,AU. The lower limit of  5\,AU corresponds to $\rho \approx 0.1''$,
well above the NICI resolution limit.

The  detection limits  in the  red  channel were  converted into  mass
ratios and  averaged over the 61  observed components, producing  a smooth
detection probability  $p_{\rm det} (r,q)$. Contours  of this function
at 0.1, 0.5, and  0.9 levels are over-plotted in Fig.~\ref{fig:q-sep}.
We see that  the chosen region of the $(r,q)$  parameter space is well
covered with NICI.

\subsection{Statistical analysis}

The purpose of this study is to reach  statistical conclusions
about the frequency and the properties of sub-systems in solar-type wide
binaries. Considering the small number of components, we keep this
analysis at a basic level, and assume that the mass ratios in
secondary sub-systems have a power-law distribution independent of
separation:
\begin{equation}
f(q) = \epsilon (\beta +1) q^\beta .
\label{eq:fq}
\end{equation}
This model has only two free parameters, the total companion frequency
$\epsilon$  and the  power-law index  $\beta$.  
Neglecting  incomplete detection, the companion
count gives $\epsilon = 7/61 = 0.12 \pm 0.04$.

The distribution of the companion  projected separations in the considered
range from 5 to 100\,AU  matters only because the detection limits
depend on $r$. We assume that $f(\log r) = {\rm const}$.

The input  data are  the $(r,q)$ values  for $K=7$ sub-systems,  the total
number  of surveyed  targets  $N =  61$,  and the  array of  detection
probabilities  $p_{\rm   det}  (r,q)$.   We  find   the  estimates  of
parameters  $(\epsilon,  \beta)$ by  the  Maximum  Likelihood (ML)  method
\citep[see  e.g.][Apendix B]{Tok06}.   The likelihood  function ${\cal
  L}$ is  a product of  the probabilities to obtain observations  for each
component  (non-detections  for $N-K$  systems  or  detections of  
$K$ subsystems with mass ratios $q_k$).

The minimum of $S = - 2 \ln {\cal L}$ is sought. This function depends
on the input data and the model parameters as 
\begin{equation}
S = 2 N f_0 - 2 \sum_{k=1}^K \ln  [ f(q_k)p_{\rm det}(r_k,q_k)] .
\label{eq:S}
\end{equation}
The summation over $k$  is done for the 7 detected companions.
The companion probability $f_0$ equals the product $f(r,q) p_{\rm
  det}(r,q)$ averaged over the considered part of the parameter space
$(r,q)$.  If the detection probability $p_{\rm det}$ equals one everywhere, the
result is $f_0 = \epsilon$, otherwise $f_0 <\epsilon$.

\begin{figure}[ht]
\epsscale{1.0}
\plotone{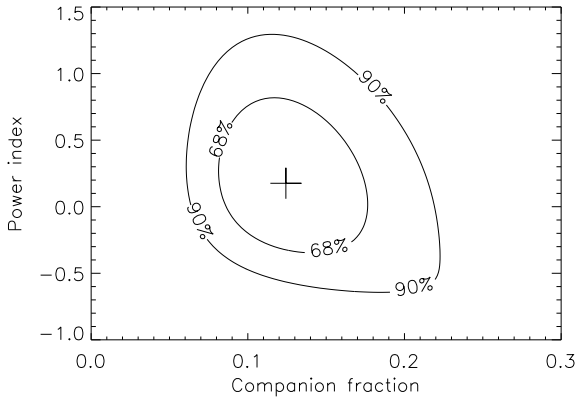}
\caption{Contours  of  the  $S$-function  in  the  $\epsilon, \beta$  space
  corresponding to the 68\% and  90\% confidence areas. The minimum is
  marked with a  cross.
\label{fig:ml1} 
}
\end{figure}

The contours of  $S$ in the parameter space  define confidence limits:
$\Delta  S =  1$ corresponds  to the  68\% interval  (``$1 \sigma$''),
$\Delta  S =  2.71$ to  90\% and  $\Delta S  = 4$  to 95\%,  in direct
analogy with the Gaussian probability distribution.  We verified the ML
method on a  simulated data set.  The input  parameters were recovered
and the confidence intervals were as expected. 

\begin{table}
\center
\caption{Confidence intervals}
\label{tab:ML}
\medskip
\begin{tabular}{c c c} 
\tableline\tableline
Interval    & $\epsilon$ & $\beta$ \\
\tableline
Min. $S$ &     0.12    &     0.18        \\
68\%     & 0.08...0.17 & $-$0.30...0.80    \\
90\%     & 0.06...0.21 & $-$0.54...1.28    \\
\tableline
\end{tabular}
\end{table}

Minimization  of (\ref{eq:S})  leads  to $(\epsilon,  \beta) =  (0.12,
0.18)$. If we ignore the detection limits and set $p_{\rm det}=1$, the
result changes  little: $(\epsilon,  \beta) = (0.11,  0.40)$.
The companion fraction 0.12 is  not different from its naive estimate.
As to the power index $\beta$, the ML points to a uniform distribution
in $q$  as being  most likely, although  the uncertainty is  large.  The
confidence areas  of the  parameters are shown  in Fig.~\ref{fig:ml1}.
Considering that the contours are not very elongated, we can determine
the 68\%  and 90\% confidence  intervals for each parameter  by a simple
cross-section through the minimum (Table~\ref{tab:ML}).
The data are  marginally compatible with $\beta \sim  -0.5$, in which case
the  fraction  of sub-systems  may  exceed  0.2  (see the  lower-right
extension of the 90\% contour in Fig.~\ref{fig:ml1}).

\section{Discussion}
\label{sec:disc}

\subsection{Completeness of the existing catalog}

The  fact that,  in a  sample of  only 33  systems, six  of  the seven
observed  companions   are  new,   shows  that  current   knowledge  of
multiplicity among  nearby solar-type systems is  very incomplete, and
that  near-IR AO  imaging is  a powerful  technique for  detecting new
companions.

\begin{figure}[ht]
\epsscale{1.0}
\plotone{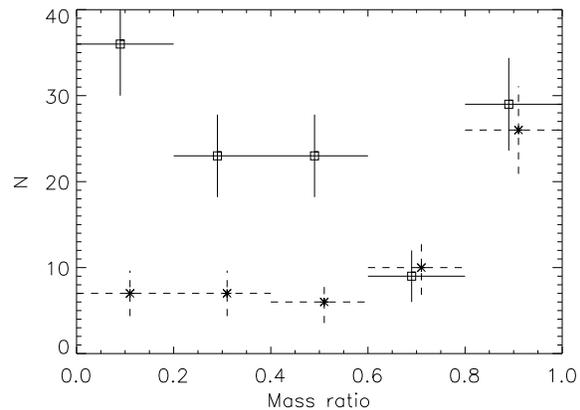}
\caption{Histograms of the mass-ratios in inner systems of triple
  stars.  Solid lines correspond to the 123 close pairs and dashed
  lines to the 61 wide pairs. The vertical bars show $\pm \sqrt{N}$
  statistical errors for each bin of width 0.2.  All points 
 are slightly displaced horizontally for clarity.
 \label{fig:q3}}
\end{figure}

We plot in Fig.~\ref{fig:q3} the mass-ratio distribution in the inner
sub-systems of known triple stars.  The data are extracted from the
updated version of the Multiple-Star Catalog  \citep[MSC,][]{MSC}, where
component masses are estimated by a variety of methods.  We selected
 220 systems within 200\,pc from the Sun with masses of primary
components from 0.5 to 1.5\,$M_\odot$.  They were sub-divided by the
separation in the inner sub-systems in two groups, 123 close
(semi-major axis $<1$\,AU) and 61 wide (between 5 and 100\,AU).  The
remaining 36 pairs have even wider separations.  The mass-ratio
distributions in these two groups are markedly different, but our
un-biased survey results in a flat mass-ratio distribution, within
errors. Therefore, the difference seen in Fig.~\ref{fig:q3} is caused
by  {\it observational selection}.  Given an approximately equal
number of close and wide sub-systems with large mass ratios (two last
bins in Fig.~\ref{fig:q3}) and assuming that the mass-ratio
distributions in these two groups are indeed similar, we expect that the
number of close and wide sub-systems (with separations below 1\,AU and
between 5 and 100\,AU, respectively) are similar, too. The
completeness of the MSC for {\it wide} sub-systems is thus about 1/2.

\subsection{Companion fraction}

Is the presence  of a wide binary companion  influencing the frequency
of inner sub-systems? To answer  this question, we compare our results
on    {\em    triples}     with    surveys    of    solar-type    {\em
  binaries}. Unfortunately, binary-star  statistics are usually derived
regardless of higher-order multiplicity,  so a cleaner comparison must
await new study of a large unbiased sample.

The separation  range from 5  to 100\,AU explored here  corresponds to
orbital periods between 10\,y  and 1000\,y. According to \citet{DM91},
about 20\% of solar-type dwarfs have companions with such periods.  In
the actual,  narrower range of  projected separations from  6.8\,AU to
44\,AU,  12\% of  solar-type dwarfs  have companions.   Therefore, the
frequency of  sub-systems around  components of wide  binaries 
 is similar to the  frequency of binary companions to single stars
in the same separation range.

The  frequency  of  sub-systems  with  periods below  3\,y  in  visual
multiples has been  determined by \citet{TS02} to be  between 11\% and
18\%.  Again, it  turned out to equal within  observational errors the
companion frequency among single dwarfs of similar masses in the solar
neighborhood and  in some open  clusters.  \citet{Desidera06} surveyed
RVs of 56  wide visual pairs and found that  $0.135 \pm 0.05$ fraction
of  companions have  spectroscopic sub-systems.   

Therefore, one  might surmise  that the presence  of a  wide companion
does not affect  the formation of inner, closer  sub-systems in a wide
range of separations.   This question is however far  from having been
settled  for both  stellar  and planetary  sub-systems \citep[see  the
  discussion  in ][]{Eggenberger07}.   We know  that almost  all close
binaries with periods below  3\,d do have additional outer companions,
although the mass  ratios of close binaries with  and without tertiary
companions have identical distributions \citep{Tok06}.

\subsection{Is the mass-ratio distribution universal?}

\citet{MH09}  advocate the  idea  of a companion  mass  function with  a
universal power law over a wide range of separations and masses.  They
fit  the mass-ratio distribution  in solar-mass  visual binaries  to a
power law with  $\beta = -0.39 \pm 0.36$.   Visual companions to stars
more massive  than the Sun follow  a power law  $f(q) \propto q^\beta$
with  $\beta$ from  $-0.3$ to  $-0.5$,  as established  in AO  imaging
surveys   by  \citet{ST02}   and   \citet{Kou05}.   Certainly,   these
distributions  do not  correspond to  the  most simplistic  case of  
random companion pairing, as described in detail by \citet{Kou09}.

Our survey of  {\em triples} points to a  flat mass-ratio distribution
with $\beta  \sim 0$.   The errors are  large, so the  difference with
visual  {\em  binaries}  is   not  yet  significant.   The  mass-ratio
distribution in  closer, spectroscopic binaries  is approximately flat
\citep{Mazeh03,Halbwachs03}, similar to the mass-ratio distribution in
  close sub-systems  of  triple stars  (Fig.~\ref{fig:q3}). {\it  If  we
confirm on a  larger sample that  sub-systems  in triples do differ
from pure binaries in  their mass-ratio distribution, its universality
will be seriously challenged.}

\subsection{Brown-dwarf desert}

The  mass function  of  companions  to binary  and  multiple stars  is
radically  different from  the Initial  Mass Function.  Therefore, the
fraction of low-mass  companions is much smaller than  the fraction of
low-mass  stars in the  field. This  fact, known  as brown  dwarf (BD)
desert,  is  well  established  \citep{Grether06}.   Adopting  a  flat
mass-ratio distribution for our sample, also typical for spectroscopic
binaries, we estimate that only 7\% of companions to a solar-mass star
will be in the BD regime.  Considering the companion frequency of 0.12
in  the  studied separation  range,  one  BD  companion per  $\sim$100
primaries is expected.  Our result thus suggests that the BD desert in
multiple systems extends to separations of 100\,AU.

The BD desert and the tendency to equal companion masses are explained
in the current star-formation scenario by continuing accretion onto a
binary.  Whenever a low-mass companion forms by fragmentation of a
massive disk, continuing accretion usually brings its mass into
stellar regime \citep[e.g.][]{Kratter09}.  Companions of sub-stellar
masses should be intrinsically rare. At the same time the orbital drag
(inward migration) reduces the separation, producing close
sub-systems.  It appears that about half of the companions settle in
close orbits with separation $<1$\,AU, the rest of them remain in
wider orbits.

\section{Conclusions}
\label{sec:concl}

We surveyed 33 nearby wide binaries with solar-type primaries and found 7
resolved sub-systems,  most of them previously unknown.  We derive the
fraction of  sub-systems with projected separations from  5 to 100\,AU
to  be   $0.12  \pm  0.04$.  The  mass-ratio   distribution  in  these
sub-systems appears to be flat, to within a large statistical error.

The sample can  be increased 6 times if  we continue observations with
NICI  and  extend  them  to  the  Northern sky  using  some  other  AO
facility. Such investment of 5 nights on 8-m telescopes will enable to
reduce the  size of the confidence intervals  in Fig.~\ref{fig:ml1} by
2.5 times.  Then, it  will become possible to distinguish between
the  negative  power index  $\beta  \sim  -0.3$  typical for  resolved
binaries   and   flat   mass-ratio   distribution   $\beta   \sim   0$
characteristic  of close  binaries.  

The  joint fraction of  both visual  and spectroscopic  sub-systems in
wide solar-type binaries is about 0.25. In consequence, the majority of the
components are single, and provide an accomodating environment to host
planetary systems.


\acknowledgments  We  thank M.   Chun  and  F.~Rigaut  for helping  us
diagnosing a NICI AO misbehaviour, and F.~Rantakyro for his assistance
in the preparation and execution  of the observations. We are thankful
for discussions on the manuscript to W.~Brandner and N.~Huelamo.  This
work used the NASA's Astrophysics  Data System, data products from the
2MASS  funded  by  the NASA  and  the  NSF,  and the  SIMBAD  database
maintained by  the University of Strasbourg, France.   The comments of
the anonymous referee helped us to improve the presentation.

{\it Facilities:} \facility{Gemini:South (NICI)}

\end{document}